\documentclass[doublecol]{epl2}

\title{All-spin logic operations: Memory device and Reconfigurable computing}

\shorttitle{All-spin logic operations}

\author{Moumita Patra\inst{1} \and Santanu K. Maiti\inst{1}\thanks{E-mail:
\email{santanu.maiti@isical.ac.in}}}

\shortauthor{Moumita Patra \em et al.}

\institute{
 \inst{1} Physics and Applied Mathematics Unit, Indian Statistical
Institute, 203 Barrackpore Trunk Road, Kolkata-700 108, India
}

 {Magnetotransport phenomena; materials for magnetotransport}
 {Spin polarized transport}

\abstract{Exploiting spin degree of freedom of electron a new proposal is given 
to characterize spin-based logical operations using a quantum interferometer 
that can be utilized as a programmable spin logic device (PSLD). The ON 
and OFF states of both inputs and outputs are described by {\em spin} state 
only, circumventing spin-to-charge conversion at every stage as often used
in conventional devices with the inclusion of extra hardware that can
eventually diminish the efficiency. All possible logic functions can be
engineered from a single device without redesigning the circuit which
certainly offers the opportunities of designing new generation spintronic
devices. Moreover we also discuss the utilization of the present model
as a memory device and suitable computing operations with proposed 
experimental setups.}

\begin{document}

\maketitle

\section{Introduction}

The use of up and down spin configurations of electron as state variables
offers new generation of spin dependent electronic devices, suppressing the 
mainstream of electronics where everything is charge based. Several 
advantages like much lower power consumption, rapid processing, 
non-volatility, higher integration densities are expected 
in the new generation spin-based systems compared to the traditional 
semi-conducting devices~\cite{s1,s2,s3,s4}. In last few years major 
attention has been paid in developing multi-functional devices involving 
electron spin such as field effect transistor, 
modulators, decoders and encoders, quantum computers to name a 
few~\cite{s1,dassarma,sp1,sp2,sp3}. The fruitful development of these 
functional devices needs much deeper understanding of proper spin dynamics.

Nowadays few ideas have been put forward to implement logical operations,
more precisely, spin-based logic gates along with the above mentioned 
functional devices. Most of the proposals for logic applications usually 
consider {\em spin state} for inputs, while output is charge 
based~\cite{LGS1,LGS2,LGS3,LGS4}. That means a spin-to-charge converter is 
required at least at one stage which definitely diminishes the overall 
advantage. However, in a recent work Dutta {\em et al.}~\cite{LGS5} have 
given a new proposal for realizing spin-based logic devices along with 
storage mechanism where spin state is considered at every stage of 
\begin{figure}[ht]
{\centering \resizebox*{6cm}{3.5cm}{\includegraphics{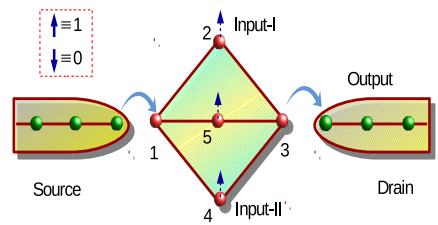}}\par}
\caption{(Color online). Sketch of all-spin logic device where a quantum 
interferometer, having five atomic sites with two non-magnetic (NM) (labeled 
as 1 and 3) and three magnetic (2, 4 and 5), is coupled to two semi-infinite 
$1$D NM source (S) and drain (D) electrodes. The orientations (up or down) 
of magnetic moments of site-2 and 4 represent the input states, and, the 
output response of the logic device is measured at the drain electrode. 
Setting the orientation (up or down) of middle magnetic site (site-5) all 
possible logical operations are obtained.}
\label{f0}
\end{figure}
operation, though
further analysis is still required to probe into it completely. Instead of
constructing individual spin logic gates it is always beneficial to design
programmable logic devices mainly because of the fact that all logic
operations can be obtained from a single device and at the same time it
reduces net cost. The another key signature is that different logical 
functions can be programmed and re-programmed to perform several 
special-purpose logic operations~\cite{LGS1,plg1,plg2,plg3} which include 
half-adder, full-adder, multiplier, spin-switches, etc. The proper 
utilization of electronic spin states in the development of memory 
technologies is another active and novel area of interest as it is supposed 
that spintronic devices are the most suitable candidates for high density 
and storage performance~\cite{LGS2,LGS5,s1,plg1,mem1}. The main effort is 
being paid to design a non-volatile cheap memory device that can offer high 
storage performance and at the same time can be accessed as fast as possible 
like random access memories (RAMs) which has not been well established so 
far to meet the present requirement of electronic industry.

Here we propose a new idea of designing all possible logical operations 
considering a single spintronic device. It can be programmed and re-programmed
for different spintronic operations and suitably engineered for data storage.
The basic device setup for achieving logical responses is illustrated in 
Fig.~\ref{f0}. The alignment of central magnetic site (site-5) plays the 
pivotal role. For one configuration a set of four different logical operations
(OR, NOR, XOR, XNOR) are obtained, while another set of four logical operations 
(AND, NAND, XOR, XNOR) are exhibited for the other configuration. The output
response is examined in terms of {\em net junction spin current} $I_s$ 
($=I_{\uparrow}-I_{\downarrow}$, where 
$I_{\sigma (\sigma=\uparrow,\downarrow)}$ being the spin dependent current),
and it is defined as high ($1$) when $I_{\uparrow}>I_{\downarrow}$ whereas
for low ($0$) output $I_{\uparrow}<I_{\downarrow}$. Qualitatively, output
response can also be determined by observing the sign of 
$T_{\uparrow}-T_{\downarrow}$ as $I_{\sigma}$ is directly involved with this 
factor where $T_{\sigma}$ denotes spin dependent transmission probability.

\section{Tight-binding Hamiltonian and theoretical prescription}

The spin dependent current ($I_{\sigma}$) through this nano-junction, 
described within a tight-binding (TB) prescription, is evaluated from 
two-terminal transmission probability ($T_{\sigma}$) which we calculate 
completely analytically. In terms of spin independent site energy $\epsilon_i$ 
($i=1$, $2$, $\dots$, $5$) and nearest-neighbor hopping (NNH) integral $t$, 
the TB Hamiltonian of the interferometer sandwiched between S and D reads as 
\begin{eqnarray}
H & = & \sum\limits_{i}c_i^{\dagger} \left(\epsilon_i-\vec{h}_i.\vec{\sigma}
\right)c_i + \sum\limits_{i} \left(c_{i+1}^{\dagger} t c_i +
h.c. \right)
\label{eq1}
\end{eqnarray}
where $c_i$, $c_i^{\dagger}$ are fermionic operators. The spin dependent
interaction term $\vec{h}_i.\vec{\sigma}$ is responsible for the separation
of up and down spin channels where $\vec{h}_i$ being the strength of magnetic
moment placed at $i$th site (non-vanishing contributions come from the sites
2, 4 and 5 only). Here we assume that $\sigma_z$ ($Z$-component of Pauli
spin matrix) is diagonal. A similar kind of TB Hamiltonian is also used 
to describe S and D, except the term $\vec{h}_i.\vec{\sigma}$ as they are 
non-magnetic, and they are parameterized by NNH strength $t_0$ and site
energy $\epsilon_0$. These electrodes are coupled to the metal channel via 
the coupling parameters $\tau_S$ and $\tau_D$. 

\begin{figure}[ht]
{\centering \resizebox*{6cm}{2cm}{\includegraphics{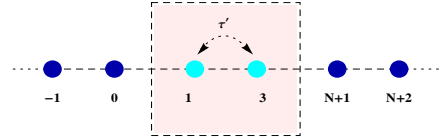}}\par}
\caption{(Color online). Renormalized 1D geometry of the quantum 
interferometer shown in Fig.~\ref{f0}, sandwiched between two metallic 
electrodes. $N$ describes the number of atomic sites in the bridging 
conductor (here $N=2$, since only two parent sites 1 and 3 are there), 
and the numbering in the electrodes can easily be followed.}
\label{rgmodel}
\end{figure}

To calculate $T_{\sigma}$ we use transfer-matrix (TM) 
method~\cite{tm1,tm2}, 
a well-known technique for studying transport properties through a 
conducting junction. First we renormalize the interferometric geometry 
(central part of Fig.~\ref{f0}) into a linear chain which effectively 
shrinks to a single bond connecting the parent atomic sites 1 and 3 as 
shown in Fig.~\ref{rgmodel}.
Under this renormalization the effective site energy ($\epsilon^{\prime}$)
of the sites 1 and 3 and hopping integral ($\tau^{\prime}$) between them 
read as (considering $\epsilon_i=0\,\forall\,i$ and $t=1\,$eV, to get a 
simple closed form structure)
\begin{eqnarray}
\epsilon^{\prime} &=& \tau^{\prime}
=\left(\begin{array}{cc}
x & 0\\
0 & y
\end{array}\right)
\end{eqnarray}
where,
\begin{eqnarray}
x&=&\frac{1}{E + h_2} + \frac{1}{E + h_4} + \frac{1}{E + h_5} \\
y&=&\frac{1}{E - h_2} + \frac{1}{E - h_4} + \frac{1}{E - h_5}
\end{eqnarray}
In the above expressions $E$ denotes the energy, and $h_2$, $h_4$ and 
$h_5$ are the strengths of magnetic moments of the magnetic sites placed 
at $2$, $4$ and $5$, respectively. From the Schr\"{o}dinger equation
$\mathcal{H}|\Phi\rangle=E |\Phi\rangle$ ($\mathcal{H}$ represents the 
Hamiltonian of the full bridge system, and 
$|\Phi\rangle=\sum_i[\psi_{i,\uparrow}|i,\uparrow\rangle + 
\psi_{i,\downarrow}|i,\downarrow\rangle]$) we can write the transfer matrix 
equation relating the wave amplitudes ($\psi_{i,\sigma}$) at sites $0$, 
$−1$ and $N + 1$, $N + 2$ as
\begin{equation}
\left(\begin{array}{cc}
    \psi_{N+2,\uparrow} \\
    \psi_{N+2,\downarrow}\\ 
    \psi_{N+1,\uparrow}\\
    \psi_{N+1,\downarrow}
\end{array}\right)
=M\left(\begin{array}{cc}
    \psi_{0,\uparrow} \\ 
    \psi_{0,\downarrow}\\
    \psi_{-1,\uparrow}\\
    \psi_{-1,\downarrow} 
\end{array}\right)
\label{equation8}
\end{equation}
where $M=M_D.P_3.P_1.M_S$ represents the transfer matrix of the complete 
system. $P_1$ and $P_3$ are the transfer matrices for the sites labeled 
as 1 and 3, respectively, and $M_S$ and $M_D$ correspond to the TMs for
the boundary sites at the left and right electrodes, respectively.
As the electrodes are perfect the wave amplitude at any particular site 
$n$ gets the form $\psi_n\sim e^{in\beta}$, where $\beta=ka$ ($a$ is
the lattice spacing). The wave vector $k$ is on the other hand related 
to the energy as $E=\epsilon_0 + 2 t_0 \cos{\beta}$. With this energy 
expression and considering the above form of wave amplitude we can write
the TMs (setting $\epsilon_0=0$, $t_0=\tau_S=\tau_D=1\,$eV) as follows:
$$M_S=M_D=\left(\begin{array}{cccc}
    e^{i\beta} & 0 & 0 & 0 \\ 
    0 & e^{i\beta} & 0 & 0 \\
    0 & 0 & e^{i\beta} & 0 \\
    0 & 0 & 0 & e^{i\beta}
\end{array}\right)$$
$$P_1=\left(\begin{array}{cccc}
    \frac{E-x}{x} & 0 & -\frac{1}{x} & 0 \\ 
    0 & \frac{E-y}{y} & 0 & -\frac{1}{y} \\
    1 & 0 & 0 & 0 \\
    0 & 1 & 0 & 0
\end{array}\right)$$
and,
$$P_3=\left(\begin{array}{cccc}
    E-x & 0 & -x & 0 \\ 
    0 & E-y & 0 & -y \\
    1 & 0 & 0 & 0 \\
    0 & 1 & 0 & 0
\end{array}\right)$$
Assuming plane wave incidence for up and down spin electrons 
Eq.~\ref{equation8} boils down to the following forms for two different 
spins:
\begin{equation}
\left(\begin{array}{cc}
    t_{\uparrow\uparrow}e^{2 i \beta} \\
    t_{\uparrow\downarrow} e^{2 i \beta}\\ 
    t_{\uparrow\uparrow}e^{i \beta} \\
    t_{\uparrow\downarrow} e^{i \beta}
\end{array}\right)
=M\left(\begin{array}{cc}
    1 + r_{\uparrow\uparrow} \\ 
    r_{\uparrow\downarrow}\\
    e^{-i\beta} + r_{\uparrow\uparrow}  e^{i\beta}\\
    r_{\uparrow\downarrow} e^{i \beta} 
\end{array}\right)
\label{eq16}
\end{equation}
and 
\begin{equation}
\left(\begin{array}{cc}
    t_{\downarrow\uparrow}e^{2 i \beta} \\
    t_{\downarrow\downarrow} e^{2 i \beta}\\ 
    t_{\downarrow\uparrow}e^{i \beta} \\
    t_{\downarrow\downarrow} e^{i \beta}
\end{array}\right)
=M\left(\begin{array}{cc}
    1 + r_{\downarrow\uparrow} \\
    r_{\downarrow\downarrow}\\
    e^{-i\beta} + r_{\downarrow\uparrow} e^{i\beta}\\
    r_{\downarrow\downarrow} e^{i \beta}
\end{array}\right)
\label{eq20}
\end{equation}
where $t_{\sigma,\sigma^{\prime}}$ and $r_{\sigma,\sigma^{\prime}}$ are
the spin-dependent transmission and reflection amplitudes. These are the 
two primary equations and solving these equations (Eqs.~\ref{eq16} and 
\ref{eq20}) we get all the required quantities. 

Doing somewhat lengthy calculations we eventually reach to the energy 
dependent up and down spin transmission probabilities as:
\begin{equation}
T_{\uparrow}(E)=T_{\uparrow\uparrow}(E)+T_{\downarrow\uparrow}(E) 
=1-\frac{(xE-1)^2}{1- 2x(E-2x)}
\label{eq2}
\end{equation}
\begin{equation}
T_{\downarrow}(E)=T_{\downarrow\downarrow}(E)+T_{\uparrow\downarrow}(E) 
=1-\frac{(yE-1)^2}{1- 2y(E-2y)}
\label{eq3}
\end{equation}
Integrating $T_{\sigma}$ over a suitable energy window, associated with 
voltage bias $V$, centering the Fermi energy $E_F$, spin dependent current 
is obtained from the relation~\cite{datta1}
\begin{equation}
I_{\sigma}(V) = \frac{e}{h} \int\limits_{E_F-\frac{eV}{2}}^{E_F+
\frac{eV}{2}}T_{\sigma}(E) \, dE
\label{eq4}
\end{equation}
Finally, net spin current is evaluated from the expression 
$I_s = I_{\uparrow} - I_{\downarrow}$.

\section{Essential results}

\begin{figure}[ht]
{\centering \resizebox*{7cm}{4cm}{\includegraphics{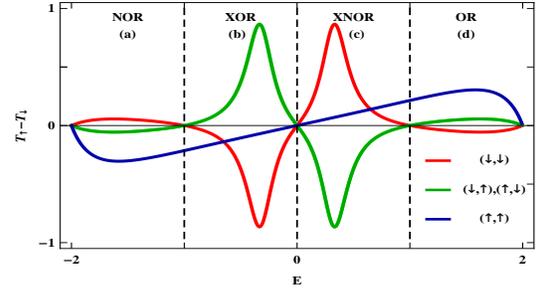}}\par}
\caption{(Color online). $T_{\uparrow}-T_{\downarrow}$ as a function of energy 
$E$ for different configurations of two inputs fixing the alignment of central 
magnetic moment (placed at site $5$) along $+$ve $Z$-direction. Four logical 
operations, NOR, NXOR, XOR, and AND, are observed since the $+$ve and $-$ve 
values of $T_{\uparrow}-T_{\downarrow}$ (upper and lower 
portions of the solid line (i.e., the line for which 
$T_{\uparrow}-T_{\downarrow}=0$) correspond to the $+$ve and $-$ve values of 
$T_{\uparrow}-T_{\downarrow}$, respectively), and hence $I_s$, represents 
the high and low output states, respectively. Here we choose $h_2=h_4=h_5=1$.}
\label{f2}
\end{figure}

Equations~\ref{eq2} and \ref{eq3} are the key expressions for analyzing 
qualitatively the all-spin logic operations in a single device.
The results are presented in Fig.~\ref{f2} for the setup where the magnetic 
moment at site $5$ is aligned along $+$ve $Z$-direction. The full allowed 
energy window ($-2\,$eV $\le E \le 2\,$eV) is divided into four
equal regions (any two such energy zones are separated by an imaginary dashed 
vertical line) where each region is associated with a distinct logical 
\begin{figure}[ht]
{\centering \resizebox*{7cm}{4cm}{\includegraphics{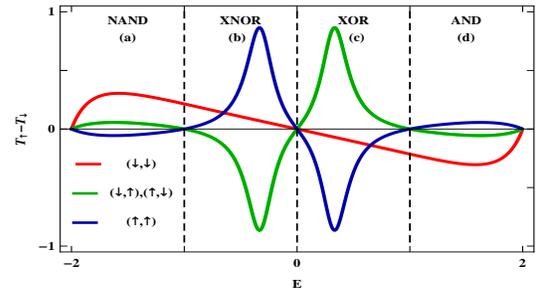}}\par}
\caption{(Color online). Same as Fig.~\ref{f2} considering the alignment
of $h_5$ along $-$ve Z-direction.}
\label{f3}
\end{figure}
operation (see Fig.~\ref{f2}). For a quantitative analysis of logical
operations i.e., in terms of net spin current $I_s$, we integrate
($T_{\uparrow}-T_{\downarrow}$) for a bias window $V=1\,$V setting the
Fermi energy $E_F$ at the centre of each distinct energy zone. The results
\begin{table}[ht]
\caption{Truth tables for different logical operations with $h_5=1$ and
$V=1\;$V. In-I and In-II describes the two inputs. $E_F$ is measured in 
unit of electron-volt.} 
$~$ 
\vskip -0.25cm
\fontsize{7}{7}
\begin{tabular}{|c|c|c|c|c|c|}
 \hline
\textbf{In-I} & \textbf{In-II} &
\multicolumn{4}{c|}{\textbf{Output ($\mbox{I}_S$ in $\mu$A)}} \\
\cline{3-6}
 &  & \textbf{NOR} &\textbf{XOR} & \textbf{XNOR} & \textbf{OR} \\
 &  & $E_F=-1.5$ & $E_F=-0.5$ & $E_F=0.5$ & $E_F=1.5$ \\
 \hline
 $\downarrow$  & $\downarrow$ & 1.5 & -11.3 & 11.3 & -1.5 \\
    $\downarrow$  & $\uparrow$  & -1.5 & 11.3 & -11.3 & 1.5 \\
    $\uparrow$  & $\downarrow$  & -1.5 & 11.3 & -11.3 & 1.5 \\
    $\uparrow$  & $\uparrow$  & -10.1 & -4.3 & 4.2 & 10.1 \\
 \hline
\end{tabular}
\label{tab1}
\end{table}
\begin{table}[ht]
\caption{Truth tables for different logical operations with $h_5=-1$ and
$V=1\;$V. The two inputs are described by In-I and In-II. $E_F$ is measured 
in unit of eV.}
$~$ 
\vskip -0.25cm
\fontsize{7}{7}
\begin{tabular}{|c|c|c|c|c|c|}
 \hline
\textbf{In-I} & \textbf{In-II} &
\multicolumn{4}{c|}{\textbf{Output ($\mbox{I}_S$ in $\mu$A)}} \\
\cline{3-6}
 &  & \textbf{NAND} &\textbf{XNOR} & \textbf{XOR} & \textbf{AND} \\
 &  & $E_F=-1.5$ & $E_F=-0.5$ & $E_F=0.5$ & $E_F=1.5$ \\
 \hline
 $\downarrow$  & $\downarrow$ & 10.1 & 4.3 & -4.3 & -10.1 \\
    $\downarrow$  & $\uparrow$  & 1.5 & -11.3 & 11.3 & -1.5 \\
    $\uparrow$  & $\downarrow$  & 1.5 & -11.3 & 11.3 & -1.5 \\
    $\uparrow$  & $\uparrow$  & -1.5 & 11.3 & -11.3 & 1.5 \\
 \hline
\end{tabular}
\label{tab2}
\end{table}
are placed in Table~\ref{tab1}. Interestingly we see that the output 
currents are reasonably high ($\sim \mu$A), and thus easy to detect.

Reversing the orientation of magnetic moment of the central magnetic site
(site $5$) we get another set of four distinct logical operations (see 
Fig.~\ref{f3}), among them two operations (XOR and XNOR) are common with 
the previous device setup. Both the qualitative (Fig.~\ref{f3}) and 
quantitative (Table~\ref{tab2}) analysis for getting different logical 
operations for this case are similar as discussed above and therefore 
we skip exploring them once again. In addition, it is important to note 
that by locking the orientation of anyone of the two inputs and altering 
the direction of the other input, NOT gate operation can easily be achieved 
as reflected from Tables~\ref{tab1} and \ref{tab2}. Thus, tuning the Fermi 
energy via suitable gate electrodes~\cite{gate1,gate2} and selectively 
orienting the central magnetic site we can in principle configure all 
possible logic gates from a single device by measuring the spin current 
at output electrode. This is one of our primary motivations behind this work.

\section{Robustness}

Now we discuss the characteristic features which 
substantiate the robustness of our proposed device in the context of logical
operations. (i) First and foremost signature is that different logical 
operations can be programmed and re-programmed by configuring the setup in
a single device. (ii) No spin-to-charge converter is required as every stage 
of operations is described by only spin states which circumvents the loss of
efficiency as usually noticed in earlier propositions. The applicability of
all spin states, on the other hand, suggests to construct all possible
functional spin-based logic functions like multiplier, half-adder, full-adder,
and to name a few. (iii) Magnitudes of output current is reasonably high
($\sim \mu$A) compared to the predicted nA currents~\cite{LGS1}, which is 
thus easy to measure more conveniently. (iv) For experimental realization 
we need to focus on two important aspects: one is spin injection efficiency
and the other is channel system i.e., whether it is semi-conducting or
a metallic one. For metal channel efficiency of spin injection is reasonably
high~\cite{LGS5}, but it exhibits less spin coherence length. Whereas for a 
semi-conducting material though coherence length is considerable, spin 
injection performance in it is extremely poor. Thus, metallic one is the
favorable option provided the issue of coherence is solved, and hopefully
it can significantly be managed by considering a smaller system. This 
is exactly what we do here, since we consider a metallic channel comprising 
only five atomic sites (too small) out of which spin dependent scattering
takes place only from three magnetic sites. (v) As the spin dependent 
scattering region is separated by NM sites in both sides of S and D, input
states are unaffected by the output states that may arise from
$\vec{\mu_i}.\vec{\mu_j}$ type interaction which yields feedback elimination
and it is an important aspect for designing efficient spintronic device.
It is also important to note that the output is determined completely by 
input states, not by any power supply. (vi) For selective switching of 
input states efficient mechanism is definitely required. One basic 
prescription of doing this is the application of local magnetic 
field~\cite{lidar,cir3}, such that it does not influence any other part of 
the device. Here we present an elegant 
\begin{figure}[ht]
{\centering \resizebox*{8.5cm}{8cm}{\includegraphics{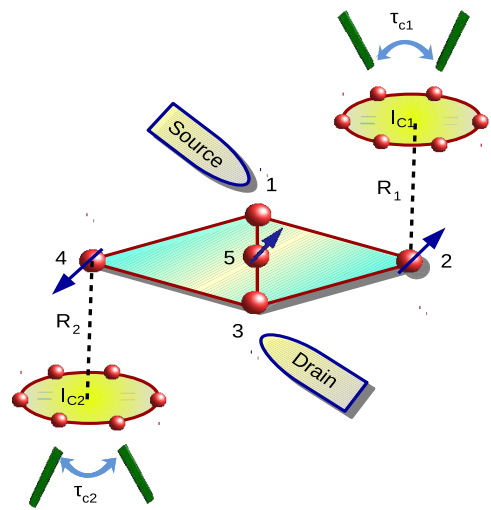}}\par}
\caption{(Color online). Possible sketch for designing all-spin logic device.
Two nano rings are placed in the vicinities of magnetic sites $2$ and $4$,
where each ring is coupled with electronic reservoirs $S_k$ and $D_k$ 
($k=1,2$). Due to close proximity of $S_k$ and $D_k$ electron can directly
hop (hopping strength is described by $\tau_k$) between them along with
the ring paths. At non-zero bias a net circular current is established in
each ring which induces a finite magnetic field. Tuning $\tau_k$ by some
mechanical means and adjusting voltage bias required magnetic field for 
specific alignment (up or down) of magnetic moment is generated.}
\label{f4}
\end{figure}
technique incorporating the idea of bias dependent circular current in a 
nano ring that induces a magnetic field at selective regions with tunable
strengths. We will discuss this mechanism elaborately in the next section.
(vii) As the size of the bridging conductor is too small energy levels are
widely separated, and thus all the logic functions can be performed even 
at moderate temperatures which is always an important issue for application
perspective. (viii) Finally, it is important to note that all the physical 
pictures studied here are observed for a wide range of bias window and remain 
invariant under a broad range of parameter values which we confirm through 
exhaustive inspections. Certainly this feature brings significant impact 
and confidence in designing an experimental setup along this line.

\section{Proposed experimental setup for logic device}

A possible experimental setup is demonstrated in Fig.~\ref{f4}. The main
concern is associated with the specific alignment of magnetic moments, 
describing the input states, by a worthy mechanism. The key concept is that 
in presence of finite bias a net circular current is established and 
\begin{figure}[ht]
{\centering \resizebox*{6cm}{3.5cm}{\includegraphics{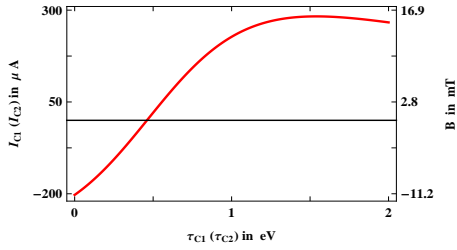}}\par}
\caption{(Color online). Dependence of $I_{ck}$ ($k=1,2$) (left axis) and
induced magnetic field $B$ ($=0,0,B$) (right axis) for a twenty-site 
ring having radius $10$\AA$\,$ at a distance $R_1=R_2=20\,$\AA$\,$ 
considering $V=0.5\,$V. Each ring is parameterized with site energy 
$\epsilon_r$ and NNH integral $t_r$, and we choose them as $\epsilon_r=0$ 
and $t_r=1\,$eV. Identical parameter values are also taken for side-attached
electrodes.}
\label{f5}
\end{figure}
because of nano sized ring the current induces a strong enough magnetic 
field~\cite{cir7,cir8} which is responsible for orienting the magnetic
moment in desired direction. Introducing a {\em shunting path} between
$S_k$ and $D_k$, and tuning the hopping strength along this
path~\cite{shunt1,shunt2} circular current, and thus magnetic field, can
be tuned in a broad range (see Fig.~\ref{f5}) along with phase reversal 
which might be very helpful 
to produce suitable magnetic field in appropriate direction. Though this 
feature (i.e., generation of local magnetic field) is quite common with 
previous works~\cite{lidar,Per,mag2}, but in those proposals the change of 
magnetic field in a wide range is no longer possible, and not such a simpler
way. In order to calculate circular current (and hence induced magnetic 
field) and to tune it by introducing shunting path in individual nano rings 
placed in the vicinities of magnetic sites 2 and 4 first we need to describe
the Hamiltonian of the system i.e., the ring with attached electrodes, and
we do it again by a tight-binding framework. Since the TB Hamiltonian of this
part and the detailed analysis of circular current along with magnetic field
are given in Ref.~\cite{tC} here we do not repeat the same part once again.
The induced magnetic field $\vec{B}$ interacts with the magnetic moment(s) 
and this term becomes $\vec{h}_{2(4)}.\vec{B}$. This is the conventional 
interaction term (called as exchange interaction) for a magnetic 
moment placed in a magnetic field. The TB Hamiltonian of the other part 
i.e., the interferometer is same as given in Eq.~\ref{eq1}.

\section{Outlines of experimental realization for memory device} 

Introducing
a free magnetic site and maintaining its magnetization direction when the
current is off we can construct non-volatile functional memory device.
\begin{figure}[ht]
{\centering \resizebox*{8cm}{8cm}{\includegraphics{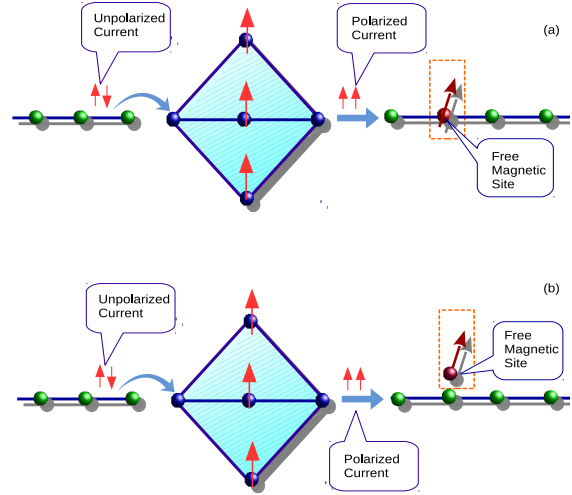}}\par}
\caption{(Color online). Two possible layouts for memory device depending 
on the location of free magnetic site. In the upper panel it is embedded 
in the wire, while in the lower panel the site is placed above the wire 
in its close proximity.}
\label{f6}
\end{figure}
Two different schemes are proposed (Fig.~\ref{f6}) depending on the
location of free magnetic site. In the upper panel (Fig.~\ref{f6}(a)) the
orientation of free site is adjusted following the mechanism of spin-transfer
torque (STT)~\cite{stt1,stt2,stt3} where spin angular momentum is 
transferred through exchange interaction, and depending on the sign of 
polarized spin current, the magnetic moment can be aligned either parallel 
or anti-parallel to the fixed magnetization direction of the metal channel. 
Similar prescription (STT) is also applicable to the other setup 
(Fig.~\ref{f6}(b)) where free magnetic site is not embedded in the drain 
wire, rather placed very close to it, as electrons can tunnel through this 
site because of close proximity along with the wired path. Another 
prescription, the so-called spin-spin exchange interaction, can also be 
implemented where direct tunneling of electrons through this site
is no longer required. For both these two setups the 
Hamiltonian of the full system can be described by incorporating an extra 
term in Eq.~\ref{eq1} which reads as 
$J\mu_{\mbox{\tiny tn}}.\mu_{\mbox{\tiny free}}$ where 
$\mu_{\mbox{\tiny tn}}$ and $\mu_{\mbox{\tiny free}}$ refer to the 
magnetic moments of the tunnel electron and free magnetic site, 
respectively, and $J$ corresponds to the interaction strength between these
magnetic moments~\cite{n1}. The main concern in the above two cases is that 
the physical mechanisms rely on net current transfer through the drain wire 
and one may think that sufficient current may not be available. But recent 
intense studies and
developments suggest that the required current density, proportional to
STT amplitude, to switch magnetization is~\cite{stt1} $\sim 10^7$ A cm$^-2$, 
and for our model it is too high mainly because of this narrow channel wire. 
Thus, higher efficiency and improved performance can be expected from our 
propositions in the development of memory based technologies.

\section{Conclusion}

We end our discussion by concluding that the proposal studied here provide
a boost in the field of storage mechanism, reconfigurable computing, 
spin-based logic functions and other several spintronic applications.

\acknowledgments

The authors gratefully acknowledge the valuable discussions with 
Prof. S. Sil and Prof. J. Ieda. MP is thankful to UGC, India 
(F. $2-10/2012$(SA-I)) for providing her doctoral fellowship.

\end{document}